\documentclass{elsart}
\usepackage{graphicx}

\begin{document}

\begin{frontmatter}
\title{Electromagnetic corrections for the analysis of 
low energy $\pi^-p$ scattering data}

\author[ZH]{A. Gashi}
\author[ZH]{E. Matsinos$^*$}
\author[AA]{G.C. Oades}
\author[ZH]{G. Rasche}
\author[CA]{W.S. Woolcock}

\address[ZH]{Institut f\"{u}r Theoretische Physik der
Universit\"{a}t,
Winterthurerstrasse 190, CH-8057 Z\"{u}rich, Switzerland}
\address[AA]{Institute of Physics and Astronomy, Aarhus University,
DK-8000 Aarhus C, Denmark}
\address[CA]{Department of Theoretical Physics, IAS,
The Australian National University, Canberra, ACT 0200, Australia}

\begin{abstract}
We calculate the electromagnetic corrections to the isospin invariant
mixing angle and to the two eigenphases for the $s$-, $p_{1/2}$- and
$p_{3/2}$-partial waves for $\pi^- p$ elastic and charge exchange scattering. 
These corrections have to be applied to the nuclear quantities in order to
obtain the two hadronic phase shifts for each partial wave. The calculation
uses relativised Schr\"{o}dinger equations containing the sum of an
electromagnetic potential and an effective hadronic potential. The mass
differences between $\pi^-$ and $\pi^0$ and between $p$ and $n$ are
taken into account. We compare our results with those of previous
calculations and estimate the uncertainties in the corrections.
\end{abstract}

$^*$ \address{Present address: The KEY Institute for Brain-Mind Research, 
University Hospital of Psychiatry, Lenggstrasse 31, CH-8029 Z\"{u}rich,
Switzerland.}
\end{frontmatter}

\section{Introduction}
 
In the previous paper \cite{1} we described the calculation of the
electromagnetic corrections to the three hadronic phase shifts with
$l=0,1$ for $\pi^+ p$ elastic scattering at pion laboratory kinetic energy
$T_{\pi} \le 100$ MeV, using relativised Schr\"{o}dinger equations (RSEs)
containing the sum of an electromagnetic potential and an effective hadronic
potential. In Section 1 of Ref.\cite{1} we gave our reasons for using such
a potential model rather than the dispersion theory model of the NORDITA
group \cite{2,3}. Our aim in this paper is to describe the corresponding
calculations for the two-channel ($\pi^- p$, $\pi^0 n$) system in the same
energy region.

The phase-shift analysis (PSA) of the low energy $\pi^+ p$ elastic scattering
data and the simultaneous calculation of the electromagnetic corrections
was possible because, ignoring the minute inelasticities due to
bremsstrahlung, only one real phase shift is needed for each partial wave. 
Such a programme is not possible for the low energy $\pi^- p$ scattering data
because three parameters (two eigenphases and a mixing angle), and therefore
three electromagnetic corrections, need to be obtained for each partial wave.
However, while the set of $\pi^- p$ elastic scattering data is only slightly
smaller (roughly 300 points) than the set of $\pi^+ p$ data, there are only
53 published data points for charge exchange scattering in our energy range.
It is quite out of the question to extract nine parameters from a PSA of
the available data without making further assumptions. The data can reliably
yield only one parameter for each partial wave, as in the $\pi^+ p$ case.

That means that we are forced in the two-channel case to invoke the
assumption of isospin invariance for the hadronic interaction in some form.
The situation is further complicated by the presence of differences between
the physical masses of the charged and neutral particles. Since we keep to the 
point of view adopted in Ref.\cite{1}, that we take account only of those
electromagnetic effects that can be calculated with reasonable confidence,
we analyse the $\pi^- p$ data and calculate the electromagnetic corrections
on the assumption that the two-channel system can be treated at the effective
hadronic level (which is certainly not the true hadronic situation when the
electromagnetic interaction is switched off) as an isospin invariant system
with all the pions having the physical mass $\mu_c$ of $\pi^{\pm}$ and both
$p$ and $n$ having the physical mass $m_p$. This implies that the hadronic
phase shifts $\delta_{l\pm}^h$, obtained from the PSA of the $\pi^+ p$ data
which went hand in hand with the calculation of the electromagnetic
corrections, are identified with phase shifts $(\delta_3^h)_{l\pm}$
corresponding to total isospin $T=3/2$. They can therefore be used as known
input to the PSA of the $\pi^- p$ data. This PSA will then yield three new
hadronic phase shifts $(\delta_1^h)_{l\pm}$ corresponding to $T=1/2$.

We therefore assume that there are real symmetric $2 \times 2$ matrices
$\mathbf{t}^h_{l\pm}$ which generalise $\tan\delta^h_{l\pm}$ from the
$\pi^+ p$ case to the two channel situation:
\begin{equation}
\mathbf{t}_{l\pm}^h=\mathbf{O}(\phi^h) \left( \begin{array}{cc}
\tan(\delta_1^h)_{l\pm} & 0 \\
0 & \tan (\delta_3^h)_{l\pm} \end{array} \right)
 \mathbf{O}(\phi^h)^t  .
\label{eq:1}
\end{equation}
The orthogonal transformation (\ref{eq:1}) takes the hadronic $t$-matrices
from the isospin basis (in which they are diagonal) to the physical basis. 
The isospin invariant mixing angle is $\phi^h=\arcsin(1/\sqrt{3})$ and 
\[
\mathbf{O}(\phi)= \left( \begin{array}{cc}
\cos \phi & \sin \phi \\
 -\sin \phi & \cos \phi \end{array} \right)  .
\]
In order to analyse the low energy $\pi^- p$ scattering data it is necessary
to calculate electromagnetic correction matrices $\mathbf{t}^{em}_{l\pm}$
which, when added to the $\mathbf{t}^h_{l\pm}$, give the real symmetric
$2 \times 2$ nuclear matrices $\mathbf{t}^n_{l\pm}$:
\begin{equation}
\mathbf{t}^n_{l\pm}=\mathbf{t}^h_{l\pm}+\mathbf{t}^{em}_{l\pm} .
\label{eq:2}
\end{equation}
The experimental observables are related to the nuclear partial wave matrices
$\mathbf{t}^n_{l\pm}$ by formulae which are given in Section 2 and in
Eqs.(1,2) of Ref.\cite{2}.

For the calculation of the $\mathbf{t}^{em}_{l\pm}$ we use the potential
model of Ref.\cite{1} in a generalisation to the two-channel case. We
emphasise that the potentials are introduced only in order to calculate the
corrections. In addition to the potentials $(V_3^h)_{l\pm}$, which we 
identify with the effective hadronic potentials $V_{l\pm}^h$ for $\pi^+p$
scattering, we have new potentials $(V_1^h)_{l\pm}$ which are constructed
in order to reproduce the phase shifts $(\delta_1^h)_{l\pm}$, using the same
RSEs containing the masses $m_p$ and $\mu_c$. We therefore have $2 \times 2$
effective hadronic potential matrices $\mathbf{V}_{l\pm}^h$ which are isospin
invariant:
\begin{equation}
\mathbf{V}_{l\pm}^h=\mathbf{O}(\phi^h) \left( \begin{array}{cc}
(V_1^h)_{l\pm} & 0 \\
0 & (V_3^h)_{l\pm} \end{array} \right)
\mathbf{O}(\phi^h)^t .
\label{eq:3}
\end{equation}
The electromagnetic correction matrices are obtained by adding electromagnetic
potential matrices $\mathbf{V}_{l\pm}^{em}$ to the $\mathbf{V}_{l\pm}^h$ and
then using these total potential matrices in coupled RSEs that model the
physical situation. Full details will be given in Section 3. The other
constraint that we impose on the effective hadronic potential matrices in
Eq. (\ref{eq:3}) is that they be {\it energy independent}. This assumption was
already made for the hadronic potentials used in the calculations for
$\pi^{+}p$ elastic scattering described in Ref. \cite{1}.
Isospin invariance identifies these as the potentials $(V^{h}_{3})_{l\pm}$
and we now require the new potentials $(V^{h}_{1})_{l\pm}$ to be energy
independent as well.

As we remarked in Ref.\cite{1}, the hadronic potential matrices in the absence
of the electromagnetic interaction would be different from the effective
matrices in its presence. Therefore isospin invariance for the
$\mathbf{V}_{l\pm}^h$ (Eq.(\ref{eq:3})) is a separate assumption. 
However, in the absence of any reliable model results for what happens when
the electromagnetic interaction is switched on, in order to analyse the
$\pi^-p$ scattering data we have no choice but to make this assumption and to
test it by seeing if it is possible to obtain a statistically acceptable fit
to the data. Implicit in this assumption is the requirement that the $T=3/2$
phase shifts for the $\pi^{-}p$ analysis be fixed at their values from 
the $\pi^{+}p$ analysis. To allow the possibility of different `$T=3/2$'
phase shifts here would already introduce the violation of isospin invariance,
which is exactly what we wish to avoid if possible. A specific model for
this violation would then be needed for a meaningful PSA to be possible, and
a modified formalism would need to be used.

We tested the assumption of isospin invariance at the effective hadronic level
and of energy independent effective hadronic potentials
by first of all analysing only the $\pi^-p$ elastic scattering data, which are
far more extensive than the charge exchange data. To anticipate results to be
given later, a PSA based on these assumptions gives a fit to the present
$\pi^{-}p$ elastic scattering data that is statistically poor but just
acceptable. If the accumulation of data eventually results in a clearly
unacceptable fit, for which there are systematic deviations of the data
from the fitted values, this would already
be evidence of `dynamical' violation of isospin invariance (that is, beyond the
effect of the electromagnetic interaction and the mass differences). The PSA
and the calculation of the electromagnetic corrections would then need to be
reconsidered. As things stand at present, the data on $\pi^{\pm}p$ elastic
scattering is {\it consistent with} the assumption of isospin invariance of
the effective hadronic interaction and of energy independent hadronic
potentials for calculating the electromagnetic corrections. This does not
provide evidence {\it for} these assumptions; an investigation of the
possibility
of dynamical violation of isospin invariance requires the study of the data
on $\pi^{-}p$ charge exchange scattering and on pionic hydrogen. Some
positive evidence from the former is given in Refs.\cite{4a} and \cite{4}.
We will
reconsider all of the evidence in a later paper concerned with a phase shift
analysis of low energy $\pi^{\pm}p$ scattering data and the comparison of the
$s$-wave scattering lengths obtained from this analysis and from the position
and width of the $1s$ level of pionic hydrogen. This paper
will complete our study of the pion-nucleon interaction at low energies.

The basic ideas sketched in this introduction will be fully developed in the
rest of the paper. Section 2 will set out the formalism for the scattering
amplitude matrices in the two-channel case, while Section 3 will give the
method of calculating the electromagnetic correction matrices for the $s$-,
$p_{1/2}$-  and $p_{3/2}$-waves. The numerical results for these corrections
will be given in Section 4.

\section{Scattering formalism}

We begin by writing the $2 \times 2$ matrices of no-flip and spin-flip
scattering amplitudes for the ($\pi^-p$, $\pi^0n$) system in the form 
\begin{equation}
{\mathbf{f}}={\mathbf{f}}^{em} + \sum_{l=0}^{\infty} \{ (l+1){\mathbf{e}}_{l+}
{\mathbf{f}}_{l+}{\mathbf{e}}_{l+}+l{\mathbf{e}}_{l-}{\mathbf{f}}_{l-}{\mathbf
{e}}_{l-} \} P_{l} ,
\label{eq:4}
\end{equation}
\begin{equation}
{\mathbf{g}}={\mathbf{g}}^{em} + i \sum_{l=1}^{\infty} ( {\mathbf{e}}_{l+}
{\mathbf{f}}_{l+}{\mathbf{e}}_{l+}-{\mathbf{e}}_{l-}{\mathbf{f}}_{l-}{\mathbf
{e}}_{l-} ) P_{l}^{1} ,
\label{eq:5}
\end{equation}
where
\begin{equation}
\mathbf{e}_{l\pm}= \left( \begin{array}{cc}
\exp(i\Sigma_{l\pm}) & 0 \\
0 & 1 \end{array} \right)  ,
\label{eq:6}
\end{equation}
\[
\Sigma_{l\pm}=(\sigma_l-\sigma_0)+\sigma_l^{ext}+\sigma_{l\pm}^{rel}+
\sigma_{l}^{vp} .
\]
The pieces of $\Sigma_{l\pm}$ are given by Eqs.(21-23) and (29) of
Ref.\cite{1}, with a change of sign in each case. Eqs.(\ref{eq:4},\ref{eq:5})
are the obvious generalisations of Eqs.(30,31) of Ref.\cite{1}. The matrices
$\mathbf{f}^{em}$, $\mathbf{g}^{em}$ are just 
\begin{equation}
\mathbf{f}^{em}= \left( \begin{array}{cc}
f^{em} & 0  \\
0 & 0 \end{array} \right)  ,  
\mathbf{g}^{em}= \left( \begin{array}{cc}
g^{em} & 0  \\
0 & 0 \end{array} \right)  , 
\label{eq:7}
\end{equation}
where the electromagnetic amplitudes  $f^{em}$, $g^{em}$ for $\pi^-p$
scattering have the same decomposition as for $\pi^+p$ scattering, namely
\begin{equation}
f^{em}=f^{pc}+f^{ext}_{1\gamma E}+f^{rel}_{1\gamma E}+f^{vp}  ,
\label{eq:8}
\end{equation}
\begin{equation}
g^{em}=g^{rel}_{1\gamma E}  .
\label{eq:9}
\end{equation} 
The expressions for the components of $f^{em}$, $g^{em}$ are those given in 
Eqs.(7)-(9), (18) and (20) of Ref.\cite{1}, with $\alpha\rightarrow-\alpha$,
$\eta\rightarrow-\eta$ in Eq.(18) and a change of sign in the other four
amplitudes. The form factors are unchanged since $F_{\pi}$ is the same for
$\pi^+$ and $\pi^-$.

The $2 \times 2$ matrices $\mathbf{f}_{l\pm}$ of partial wave amplitudes are
written most conveniently in the form 
\begin{equation}
\mathbf{f}_{l\pm}=  \left( \begin{array}{cc}
q_c^{-1/2} & 0 \\
0 & q_0^{-1/2}  \end{array} \right) \mathbf{T}_{l\pm}^n  
\left( \begin{array}{cc}
q_c^{-1/2} & 0 \\
0 & q_0^{-1/2}  \end{array} \right)   ,
\label{eq:10}
\end{equation}
\begin{equation}
\mathbf{T}_{l\pm}^n=\mathbf{t}_{l\pm}^n(\mathbf{1}_2-i\mathbf{t}_{l\pm}^n)^
{-1}  ,
\label{eq:11}
\end{equation}
where $q_c$ is given by Eq.(10) of Ref.\cite{1} and $q_0$ is the corresponding
c.m. momentum for the $\pi^0n$ channel,
\begin{equation}
q_0^2=\frac{[W^2-(m_n-\mu_0)^2][W^2-(m_n+\mu_0)^2]}{4W^2}  ,
\label{eq:12}
\end{equation}
$m_n$ and $\mu_0$ being the masses of the neutron and $\pi^0$ respectively. 
The nuclear matrices $\mathbf{t}_{l\pm}^n$ in Eq.(\ref{eq:11}) are those
introduced in Section 1. 

The expressions (\ref{eq:10}) and (\ref{eq:11}), with real symmetric matrices
$\mathbf{t}_{l\pm}^n$, assume two-channel unitarity, that is the absence of
any competing channels. This is not quite true, since the $\gamma n$ channel
introduces inelastic corrections to the $\mathbf{T}_{l\pm}^n$ which need to
be taken into account. In the energy range $T_{\pi}\leq 100$ MeV these
corrections are almost insignificant and it is sufficient to use the results
of Ref.\cite{2}, which are derived from known amplitudes for the reactions
$\gamma n \rightarrow \pi^-p$, $\pi^0 n$ using three-channel unitarity. It is
unnecessary to introduce a complex hadronic potential matrix. 
The observables for $\pi^-p$ elastic and charge exchange scattering need to
be calculated from the partial wave amplitudes $f_{cc}$, $f_{0c}$ given by
Eq.(\ref{eq:10}), with $T^n_{cc}$, $T^n_{0c}$ replaced by $T_{cc}$, $T_{0c}$,
where 
\[
T_{cc}=T^n_{cc}+\Delta T_{cc}^{\gamma n} ,   T_{0c}=T^n_{0c}+
\Delta T_{0c}^{\gamma n} .
\]
We have dropped the subscript $l\pm$ for convenience and have used the
subscripts $c$,$0$ to denote the channels $\pi^-p$, $\pi^0 n$ respectively. 
The nuclear quantities $T^n_{cc}$, $T^n_{0c}$ are calculated from
Eq.(\ref{eq:11}) with the real symmetric matrix $\mathbf{t}^n$ given by 
Eq.(\ref{eq:1}). The corrections $\Delta T_{cc}^{\gamma n}$, $\Delta T_{0c}^
{\gamma n}$ (for the $s$- and $p_{3/2}$-waves) are related to the quantities 
$\overline{\eta}_1$, $\overline{\eta}_3$ and $\eta_{13}$ in Table IV of
Ref.\cite{2} by the formulae
\begin{equation}
2i\Delta T_{cc}^{\gamma n}=-\frac{2}{3}\overline{\eta}_1\exp(2i\delta_1^h)-
\frac{1}{3}\overline{\eta}_3\exp(2i\delta_3^h)-\frac{8}{9}{\eta}_{13}\exp
\{i(\delta_1^h+\delta_3^h)\}  ,
\label{eq:13}
\end{equation}
\begin{equation}
2i\Delta T_{0c}^{\gamma n}=\frac{\sqrt{2}}{3}\overline{\eta}_1\exp(2i\delta_1
^h)-\frac{\sqrt{2}}{3}\overline{\eta}_3\exp(2i\delta_3^h)-\frac{2\sqrt{2}}{9}
{\eta}_{13}\exp\{i(\delta_1^h+\delta_3^h)\}  .
\label{eq:14}
\end{equation}

The decomposition of the matrices $\mathbf{t}_{l\pm}^n$ into a hadronic part
$\mathbf{t}_{l\pm}^h$ and its electromagnetic correction $\mathbf{t}_{l\pm}^
{em}$ is given in Eq.(\ref{eq:2}). As we have discussed, the matrices 
$\mathbf{t}_{l\pm}^h$ refer to an effective hadronic situation which is
isospin invariant and in which all the pions have the mass $\mu_c$ and the
nucleons the mass $m_p$. The aim of our calculation is to obtain for the
$s$-, $p_{1/2}$- and $p_{3/2}$-waves the three independent elements of the
symmetric correction matrix $\mathbf{t}_{l\pm}^{em}$. However, the
corrections in this form do not convey information in the way we would
naturally like to have it. It is therefore customary to write the nuclear
matrices in the form 
\begin{equation}
\mathbf{t}^n=\mathbf{O}(\phi) \left( \begin{array}{cc}
\tan \delta_1^n & 0 \\
0 & \tan \delta_3^n \end{array} \right) 
\mathbf{O}(\phi)^t
\label{eq:15}
\end{equation}
and to define three new corrections $C_1$, $C_3$ and $\Delta \phi$ by 
\begin{equation}
C_1=\delta_1^n-\delta_1^h  ,  C_3=\delta_3^n-\delta_3^h  ,   \Delta
\phi=\phi-\phi^h  .
\label{eq:16}
\end{equation}
Here $\phi$ is the mixing angle, which we choose to lie between $0$ and
$\pi/2$. This convention then fixes the labelling of the eigenphases and
ensures that $\tan \delta_i^n$ is close to $\tan \delta_i^h$ ($i=1,3$).

To proceed with the calculation of the corrections we need the potential
matrices ${\mathbf{V}}^h_{l\pm}$ and ${\mathbf{V}}^{em}_{l\pm}$ which appear
in the coupled RSEs that lead to the nuclear matrices $\mathbf{t}_{l\pm}^{n}$.
The effective hadronic potential matrices have the isospin invariant form of
Eq.(\ref{eq:3}) and the potentials $V_{\alpha}^h$ ($\alpha=1,3$) are
constructed so as to reproduce the hadronic phase shifts $\delta_{\alpha}^h$
via the RSEs
\begin{equation}
\left (\frac{d^2}{dr^2}-\frac{l(l+1)}{r^2}+q_c^2-2m_cf_{c}V_{\alpha}^h(r)
\right ) u_{\alpha}(r)=0  .
\label{eq:17}
\end{equation}
The phase shifts are given by the asymptotic behaviour $\sin(q_cr-l\pi/2+
\delta_{\alpha}^h)$ of the regular wavefunctions. The quantities $q_c$ and
$f_c$ are defined in Eq.(10) of Ref.\cite{1} and $m_c$ is the reduced mass of
the $\pi^- p$ system. The electromagnetic potential matrix ${\mathbf{V}}^{em}
_{l\pm}$ is 
\begin{equation}
{\mathbf{V}}^{em}_{l\pm}= \left( \begin{array}{cc}
V^{em}_{l\pm} & 0 \\
0 & 0 \end{array} \right)
\label{eq:18}
\end{equation}
and $V^{em}_{l\pm}$ has the decomposition given in Eq.(25) of Ref.\cite{1}:
\begin{equation}
V^{em}_{l\pm}=V^{pc}+V^{ext}+V^{rel}_{l\pm}+V^{vp}  .
\label{eq:19}
\end{equation}
The full potential matrix is
\begin{equation}
\mathbf{V}_{l\pm}={\mathbf{V}}_{l\pm}^{em}+\mathbf{V}^h_{l\pm}  .
\label{eq:20}
\end{equation}
The pieces of $V^{em}_{l\pm}$ in Eq.(\ref{eq:17}), and therefore $V^{em}_
{l\pm}$ itself, have the opposite sign compared with the case of $\pi^+p$
scattering. (This is true for $V^{rel}_{l\pm}$ since $\sigma^{rel}_{l\pm}$
is calculated only to order $\alpha$.) Full details of the parts of $V^{em}
_{l\pm}$ were given in Section 2 of Ref.\cite{1}.

\section{Evaluation of the corrections}

The partial wave RSEs for the two-channel case, which we use in order to
model the physical situation, are given by the natural generalisation of
Eq.(34) of Ref.\cite{1}:
\begin{equation}
\left \{(\frac{d^2}{dr^2}-\frac{l(l+1)}{r^2})\mathbf{1}_2+\mathbf{Q}^2-2
\mathbf{m}\mathbf{f}\mathbf{V}_{l\pm}(r) \right \} \mathbf{u}_{l\pm}(r)=
\mathbf{0} .
\label{eq:21}
\end{equation}
The full potential matrices $\mathbf{V}_{l\pm}$ have the form given in
Eqs.(\ref{eq:3},\ref{eq:18}-\ref{eq:20}). The matrices $\mathbf{Q}$,
$\mathbf{m}$ and $\mathbf{f}$ are 
\begin{equation}
\mathbf{Q}= \left( \begin{array}{cc}
q_c & 0  \\
0 & q_0 \end{array} \right)  ,  
\mathbf{m}= \left( \begin{array}{cc}
m_c & 0  \\
0 & m_0 \end{array} \right)  , 
\mathbf{f}= \left( \begin{array}{cc}
f_c & 0  \\
0 & f_0 \end{array} \right)  ,
\label{eq:22}
\end{equation}
where $q_0$ is defined in Eq.(11) and 
\begin{equation}
m_0=\frac{m_n\mu_0}{m_n+\mu_0}  ,  f_0=\frac{W^{2}-m_n^2-\mu_0^2}{2m_0W}
 .
\label{eq:23}
\end{equation}

The only nonzero entry of the electromagnetic potential term $2\mathbf{m}
\mathbf{f}\mathbf{V}^{em}$ is just $2m_cf_cV^{em}$, which is the same as for
$\pi^+p$ but with the change of sign in $V^{em}$. The particular form 
$2\mathbf{m}\mathbf{f}\mathbf{V}^{h}$ of the hadronic potential term, with
$\mathbf{V}^{h}$ a real symmetric energy independent matrix satisfying
isospin invariance, is crucial for the calculation of the $\pi^-p$
electromagnetic corrections and we need to explain in detail why we chose it
in this way. The issue is how to incorporate the physical mass differences
into the two-channel problem. There is no guidance from the chain of reasoning
that starts from the Bethe-Salpeter equation and proceeds via a 
three-dimensional reduction to integral equations for the partial wave
amplitudes in momentum space, which contain a hadronic quasipotential. 
This method has been used for example in Ref.\cite{5} to develop a dynamical
model for low energy pion-nucleon scattering. Such a procedure ignores the
mass differences and involves delicate issues like the choice of the reduction
and the form factors at the vertices of the tree diagrams used. The partial
wave quasipotentials, if converted to coordinate space, are nonlocal and
energy dependent. Trying to incorporate the electromagnetic interaction and
the mass differences into such a model in order to calculate the
electromagnetic corrections would be impossible. 

Since we are dealing with a spin $0$- spin $1/2$ system with the fermion much
heavier than the boson, we are not far from the static limit and it is natural
to look to the Klein-Gordon equation for guidance. For this equation with the
potential $V^{pc}$ there is a simple transformation to an RSE with
$\mu_c$ replaced by $(\mu_c^2+q_c^2)^{1/2}$, the static limit of 
$m_cf_c=(W^2-m_p^2-\mu_c^2)/2W$. The Appendix of Ref.\cite{6} assumes that
the hadronic potential enters the Klein-Gordon equation in the same way 
as $V^{pc}$, as the timelike component of a four-vector. It is then shown
that the effective hadronic potential $V^{h}$ that appears in the resulting
RSE must also be multiplied by $(\mu_c^2+q_c^2)^{1/2}$. However, it is also
possible to introduce the hadronic potential as a scalar or as a combination
of the two, so the assumption of Ref.\cite{6}, which is also made in the model
of low energy pion-nucleon scattering of Ref.\cite{7}, requires further 
justification. Looking at $\pi^+p$ alone cannot decide anything, as we said
in Ref.\cite{1}, but there is empirical evidence from the ($\pi^-p$, $\pi^0n$)
system for the choice made in Refs.\cite{6,7}.

At the hadronic level, with mass differences included, a study of the
extrapolation of the invariant amplitudes to the Cheng-Dashen point favours
this choice \cite{8}; this provides some justification for the model used in
Ref.\cite{7}. As shown in Ref.\cite{6}, this implies that the hadronic
potential term should have the form $2\mathbf{m}\mathbf{f}\mathbf{V}^{h}$ 
given in Eq.(\ref{eq:21}). Writing this term in full is instructive; it is
\[
\left( \begin{array}{cc}
2m_cf_c(\frac{2}{3}V_1^h+\frac{1}{3}V_3^h) & 2m_cf_c\frac{\sqrt{2}}{3}(V_3^h-
V_1^h) \\
2m_0f_0\frac{\sqrt{2}}{3}(V_3^h-V_1^h) & 2m_0f_0(\frac{1}{3}V_1^h+\frac{2}{3}
V_3^h) \end{array} \right)  .
\]
It is then clear that introducing the energy dependent factors $f_c$, $f_0$
is {\it not} equivalent to having nonrelativistic reduced masses $m_c$,
$m_0$ and energy dependent hadronic potentials for each isospin. The effect
of the factors $f_c$, $f_0$ is large yet subtle; they change the way in which
the violation of isospin invariance due to the mass differences is introduced.
We studied the effect of choosing the hadronic potential term as $2\mathbf{m}
\tilde{\mathbf{V}}^{h}$, {\it without} the factor $\mathbf{f}$ and with
$\tilde{\mathbf{V}}^{h}$ energy independent and isospin invariant. However,
this resulted in large changes to some of the electromagnetic corrections.
The most dramatic effect was on $C_{3}$ for the $p_{3/2}$-wave, which became
much smaller; it changed sign near 70 MeV and was $-0.4^{\circ}$ at 100 MeV,
compared with $+0.5^{\circ}$ obtained with the choice $2\mathbf{m}\mathbf{f}
\mathbf{V}^{h}$ of the hadronic potential term in Eq.(\ref{eq:21}).
The NORDITA value \cite{2} at this energy is $+0.85^{\circ}$. The correction
given by the potential model without the factor $\mathbf{f}$ is therefore
completely at variance with that given by NORDITA. Their calculation did take
account of the mass differences, though it is impossible to recover from their
references exactly how this was done.

With the electromagnetic corrections calculated using the factor $\mathbf{f}$
in the hadronic potential term, the value of $\chi^2$ for the fit to 224 data
points for $\pi^{-}p$ elastic scattering is 471.0. When the factor $\mathbf{f}
$ is absent, $\chi^{2}$ increases to 484.8, so the fit becomes worse. Most of
the increase comes from the data near 100 MeV, where the value of $C_{3}$ for
the $p_{3/2}$-wave changes so much when the factor $\mathbf{f}$ is absent. By
the usual statistical criteria, both fits
are extremely poor, so once again decisive evidence is elusive. The large
values of $\chi^{2}$ come from the data base itself. For the fit with the
factor $\mathbf{f}$ included,
there is no evidence of any systematic deviation, with either angle or energy,
of the data points from the fitted curves. The deviations are erratic and it
seems clear that in many cases the errors, particularly the systematic
errors, have been
underestimated. In the sense of giving a reasonable averaging over an
internally inconsistent body of data, the fit with the factor $\mathbf{f}$
included is certainly better than that where this factor is absent.
In summary, we have given three pieces of evidence that favour the inclusion
of the specific energy dependence introduced by the factor $\mathbf{f}$ in
the hadronic potential term: the extrapolation of the invariant amplitudes
to the Cheng-Dashen point, the comparison with the results of NORDITA \cite{2}
(who also include the mass differences) and the better fit to the data (judged
not only by the value of $\chi^{2}$ but also by the systematic deviation
near 100
MeV when $\mathbf{f}$ is not included).     
 
We turn now to some calculational details. The RSEs (\ref{eq:21}) are
integrated outwards from $r=0$ to obtain two linearly independent regular
solution vectors.  The integration proceeds to a distance $R$ (around 1000 fm)
where the only part of the potential matrix that is not negligible is
$V^{pc}$, which appears in $V_{cc}$. The components $V_{0c}$ and $V_{00}$ of
$\mathbf{V}$, which contain only linear combinations of the hadronic potential
, become negligible beyond distances of a few fm, so the integration as far
as $r=R$ is necessary only for the first component of $\mathbf{u}_{l\pm}$.

The matching at $r=R$, which leads to the matrix $\mathbf{t}_{l\pm}^n$,
requires particular attention. In the $\pi^-p$ channel the matching is to a
linear combination of the standard point charge Coulomb wavefunctions
$F_l(-\eta f_c;q_cr)$ and $G_l(-\eta f_c;q_cr)$ and in the $\pi^0 n$ channel it
is to a linear combination of the free particle wavefunctions $q_0rj_l(q_0r)$ 
and $q_0rn_l(q_0r)$. To make the notation less complicated, we drop the
subscripts $l\pm$ and $l$ for the moment and form  the matrix $\mathbf{u}(r)$:
\begin{equation}
\mathbf{u}(r)= \left( \begin{array}{cc}
(\mathbf{u}^{(1)}(r))_c & (\mathbf{u}^{(2)}(r))_c \\
(\mathbf{u}^{(1)}(r))_0 & (\mathbf{u}^{(2)}(r))_0 \end{array} \right)  ,
\label{eq:24}
\end{equation}
where $\mathbf{u}^{(i)}$, $i=1,2$, are two linearly independent solution
vectors. For $r>R$, $\mathbf{u}(r)$ has the form 
\begin{equation}
\mathbf{u}(r)=\mathbf{m}^{1/2}\mathbf{f}^{1/2}\mathbf{Q}^{-1/2}(\hat
{\mathbf{f}}(r)\mathbf{a}+\hat{\mathbf{g}}(r)\mathbf{b}) ,
\label{eq:25}
\end{equation}
where
\begin{equation}
\hat{\mathbf{f}}(r)= \left( \begin{array}{cc}
\hat{f}_c(r) & 0 \\
0 & \hat{f}_0(r) \end{array} \right)  , 
\hat{\mathbf{g}}(r)= \left( \begin{array}{cc}
\hat{g}_c(r) & 0  \\
0 & \hat{g}_0(r) \end{array} \right)  ,
\label{eq:26}
\end{equation}
\begin{equation}
\hat{f}_c(r)=\cos(\Delta\sigma)F(-\eta f_c;q_cr)+\sin(\Delta\sigma)
G(-\eta f_c;q_cr) ,
\label{eq:27}
\end{equation}
\begin{equation}
\hat{g}_c(r)=\cos(\Delta\sigma)G(-\eta f_c;q_cr)-\sin(\Delta\sigma)
F(-\eta f_c;q_cr) ,
\label{eq:28}
\end{equation}
\begin{equation}
\hat{f}_0(r)=rj(q_0r) , \hat{g}_0(r)=rn(q_0r).
\label{eq:29}
\end{equation}

The particular forms of $\hat{f}_c(r)$, $\hat{g}_c(r)$ in Eqs.(\ref{eq:27},
\ref{eq:28}) arise from the choice of the additive electromagnetic amplitudes
$(f,g)^{em}$, which contain the parts $(f,g)^{ext}$, $(f,g)^{rel}$ and 
$(f,g)^{vp}$, and of the electromagnetic phase shifts $\Sigma _{l\pm}$, which
contain $\Delta \sigma_{l\pm}=\sigma^{ext}_l+\sigma^{rel}_{l\pm}+\sigma^
{vp}_l$. This means that, in order to obtain the correct matrices
$\mathbf{t}_{l\pm}^n$ as defined in Eq.(\ref{eq:10}), it is necessary to match
to the linear combinations of the point charge Coulomb wavefunctions given in
Eqs.(\ref{eq:27},\ref{eq:28}), which correspond to the phase shift
$\sigma_l+\Delta \sigma_{l\pm}$. The nuclear matrices $\mathbf{t}^n_{l\pm}$
are given by
\begin{equation}
\mathbf{t}^n_{l\pm}=\mathbf{b}_{l\pm}\mathbf{a}_{l\pm}^{-1}  .
\label{eq:30}
\end{equation}
The factor $\mathbf{m}^{1/2}\mathbf{f}^{1/2}\mathbf{Q}^{-1/2}$ in
Eq.(\ref{eq:25}) appears naturally if one works with a symmetric matrix 
$\mathbf{t}^n$, as shown in Section 3 of Ref.\cite{10}. Eq.(\ref{eq:30}) is
the generalisation to two channels of the one-channel result given in Eq.(40)
of Ref.\cite{1}. 

\begin{figure}
\begin{center}
\includegraphics[height=0.55\textheight,angle=0]{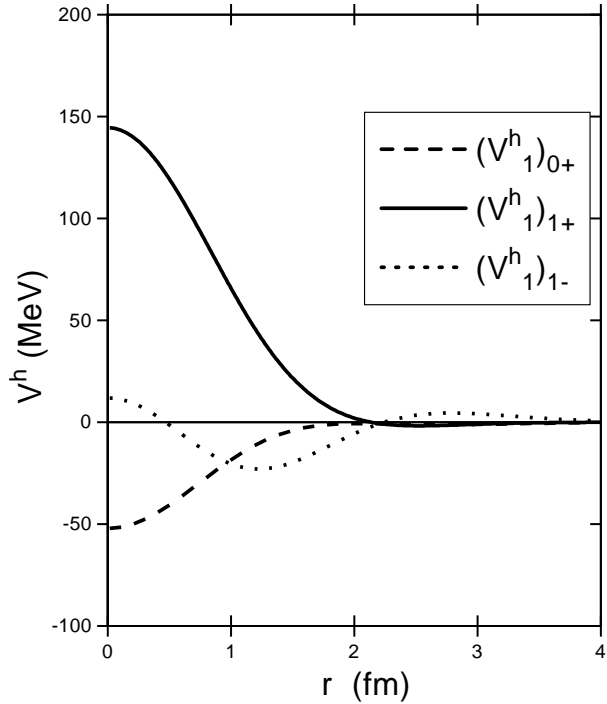}
\caption{The $T=1/2$ hadronic potentials $(V^h_1)_{0+}$ and  $(V^h_1)_{1\pm}$.}
\label{fig:1}
\end{center}
\end{figure}

\begin{table}
\begin{center}
\caption
{Values in degrees of the $s$-wave electromagnetic corrections $C_{3}$, $C_{1}$
and $\Delta \phi$ as functions of the pion lab kinetic energy $T_{\pi}$
(in MeV).}
\label{tab:1}
\begin{tabular}{|c|c|c|c|}
\hline
 $T_{\pi}$ & $C_{3}$ & $C_{1}$ &$\Delta \phi$ \\\hline
 10&-0.199$\pm$ 0.007&0.208$\pm$ 0.002&0.274$\pm$ 0.013\\
 15&-0.175$\pm$ 0.008&0.163$\pm$ 0.002&0.116$\pm$ 0.011\\
 20&-0.161$\pm$ 0.009&0.133$\pm$ 0.001&0.039$\pm$ 0.011\\
 25&-0.151$\pm$ 0.010&0.111$\pm$ 0.001&-0.002$\pm$ 0.010\\
 30&-0.145$\pm$ 0.010&0.093$\pm$ 0.001&-0.025$\pm$ 0.010\\
 35&-0.140$\pm$ 0.011&0.080$\pm$ 0.002&-0.038$\pm$ 0.010\\
 40&-0.136$\pm$ 0.011&0.069$\pm$ 0.002&-0.044$\pm$ 0.010\\
 45&-0.134$\pm$ 0.011&0.060$\pm$ 0.003&-0.046$\pm$ 0.009\\
 50&-0.132$\pm$ 0.012&0.052$\pm$ 0.003&-0.045$\pm$ 0.009\\
 55&-0.130$\pm$ 0.012&0.046$\pm$ 0.003&-0.043$\pm$ 0.009\\
 60&-0.128$\pm$ 0.013&0.041$\pm$ 0.003&-0.040$\pm$ 0.008\\
 65&-0.127$\pm$ 0.014&0.037$\pm$ 0.003&-0.036$\pm$ 0.008\\
 70&-0.126$\pm$ 0.015&0.033$\pm$ 0.002&-0.032$\pm$ 0.007\\
 75&-0.124$\pm$ 0.015&0.031$\pm$ 0.002&-0.027$\pm$ 0.006\\
 80&-0.123$\pm$ 0.016&0.028$\pm$ 0.002&-0.023$\pm$ 0.005\\
 85&-0.122$\pm$ 0.018&0.026$\pm$ 0.003&-0.019$\pm$ 0.005\\
 90&-0.120$\pm$ 0.019&0.024$\pm$ 0.005&-0.015$\pm$ 0.004\\
 95&-0.118$\pm$ 0.020&0.022$\pm$ 0.006&-0.011$\pm$ 0.003\\
100&-0.117$\pm$ 0.021&0.021$\pm$ 0.007&-0.007$\pm$ 0.002\\ \hline
\end{tabular}
\end{center}
\end{table}

\begin{figure}
\begin{center}
\includegraphics[height=0.55\textheight,angle=0]{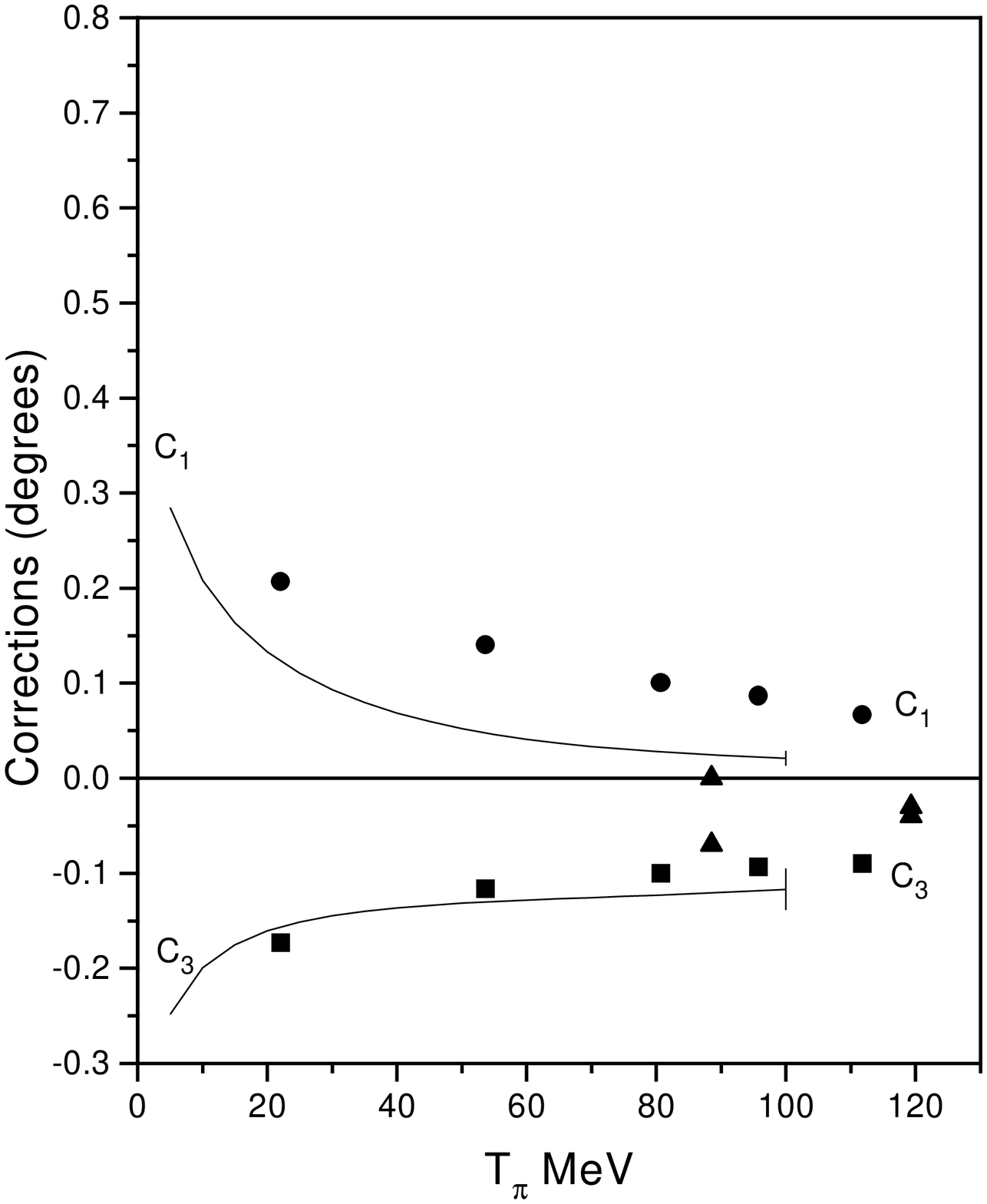}
\caption{Values in degrees of the electromagnetic corrections $C_{1}$ and
$C_{3}$ for the $s$-wave from our present calculation (solid curves), from
NORDITA \cite{2} (circles) and from Zimmermann \cite{9} (triangles).}
\label{fig:2}
\end{center}
\end{figure}

\begin{figure}
\begin{center}
\includegraphics[height=0.55\textheight,angle=0]{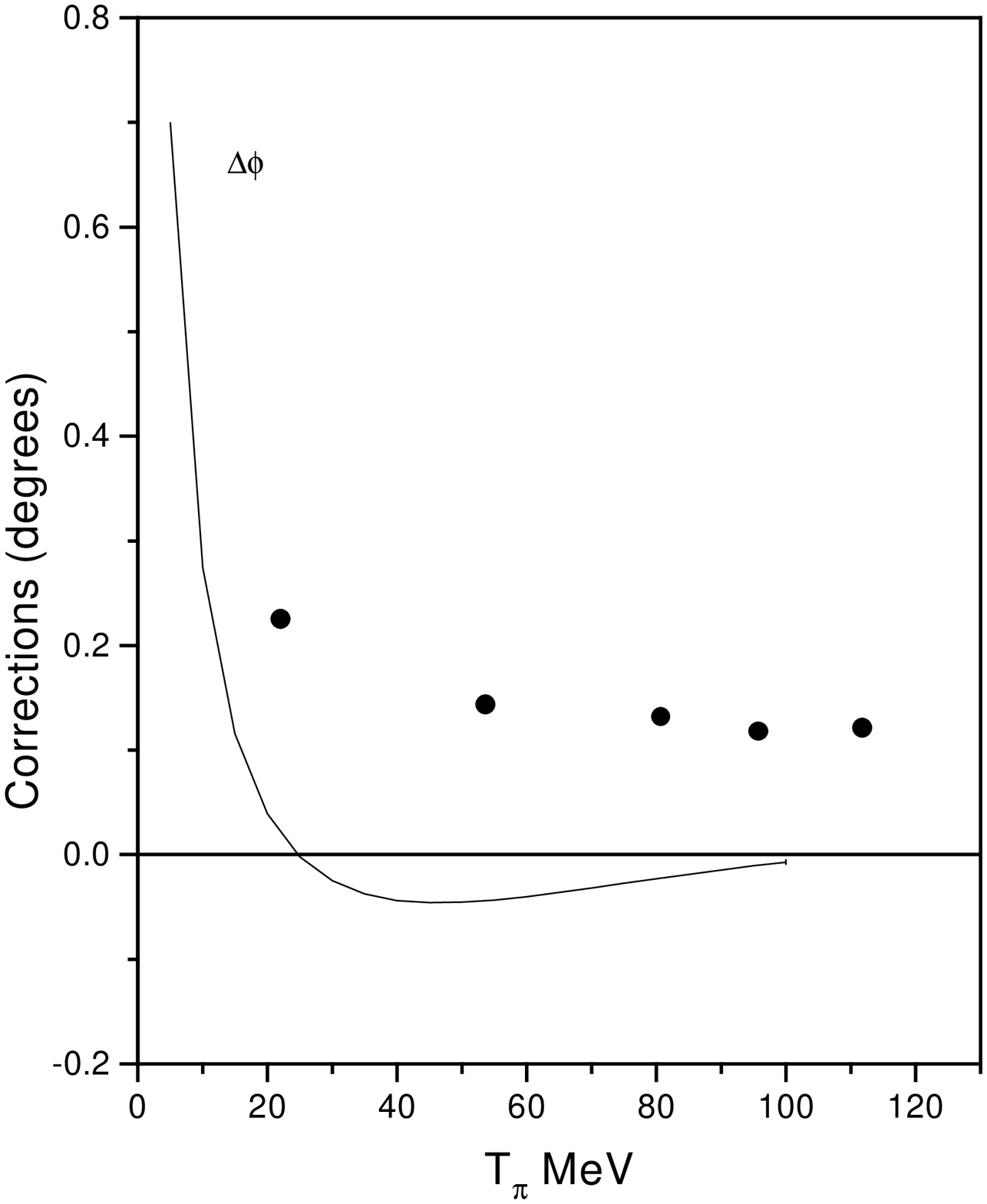}
\caption{Values in degrees of the electromagnetic correction $\Delta \phi$
for the $s$-wave from our present calculation (solid curve), from
NORDITA \cite{2} (circles) and from Zimmermann \cite{9} (triangles).}
\label{fig:3}
\end{center}
\end{figure}

The machinery for an iterative procedure exactly like that described in
Ref.\cite{1} for the $\pi^+p$ case has now been fully explained. The details
of the PSA of the $\pi^-p$ elastic scattering data will be given in a separate
paper. The hadronic phase shifts $\delta_3^h$ were fixed throughout the
$\pi^-p$ PSA at the final values from the $\pi^+p$ PSA. The starting point for
the $T=1/2$ hadronic phase shifts was the values from the analysis of Arndt
et al. \cite{11}. The parametric form of the $T=1/2$ hadronic potentials was
taken to be the same as that used for the hadronic potentials in Ref.\cite{1}. 
In Fig.\ref{fig:1} we show these potentials for the final step of the
iteration. At each step of the iteration the matrices 
$\mathbf{t}_{0+}^{em}$, $\mathbf{t}_{1-}^{em}$ and $\mathbf{t}_{1+}^{em}$
were calculated using Eqs.(\ref{eq:30}),(\ref{eq:1}) and (\ref{eq:2}). 
The conversion to the corrections in the form $C_1$, $C_3$ and $\Delta
\phi$ defined in Eq.(\ref{eq:16}) was done after the final values of these
three matrices had been obtained. 

\begin{table}
\begin{center}
\caption
{Values in degrees of the $p_{3/2}$-wave electromagnetic corrections $C_{3}$,
$C_{1}$ and $\Delta \phi$ as functions of the pion lab kinetic energy
$T_{\pi}$ (in MeV).}
\label{tab:2}
\begin{tabular}{|c|c|c|c|}
\hline
 $T_{\pi}$ & $C_{3}$ & $C_{1}$ &$\Delta \phi$ \\\hline
 10&0.159$\pm$ 0.001&-0.002$\pm$ 0.000&-4.715$\pm$ 0.049\\
 15&0.187$\pm$ 0.001&-0.003$\pm$ 0.000&-3.044$\pm$ 0.038\\
 20&0.209$\pm$ 0.002&-0.005$\pm$ 0.000&-2.146$\pm$ 0.030\\
 25&0.229$\pm$ 0.003&-0.007$\pm$ 0.000&-1.592$\pm$ 0.025\\
 30&0.246$\pm$ 0.003&-0.009$\pm$ 0.000&-1.222$\pm$ 0.020\\
 35&0.261$\pm$ 0.003&-0.011$\pm$ 0.000&-0.961$\pm$ 0.015\\
 40&0.276$\pm$ 0.002&-0.013$\pm$ 0.000&-0.769$\pm$ 0.009\\
 45&0.290$\pm$ 0.002&-0.016$\pm$ 0.000&-0.623$\pm$ 0.010\\
 50&0.305$\pm$ 0.005&-0.018$\pm$ 0.001&-0.511$\pm$ 0.014\\
 55&0.320$\pm$ 0.009&-0.021$\pm$ 0.001&-0.422$\pm$ 0.018\\
 60&0.336$\pm$ 0.012&-0.023$\pm$ 0.001&-0.351$\pm$ 0.021\\
 65&0.352$\pm$ 0.015&-0.026$\pm$ 0.001&-0.293$\pm$ 0.023\\
 70&0.370$\pm$ 0.019&-0.029$\pm$ 0.001&-0.246$\pm$ 0.024\\
 75&0.390$\pm$ 0.023&-0.031$\pm$ 0.001&-0.207$\pm$ 0.024\\
 80&0.410$\pm$ 0.026&-0.034$\pm$ 0.001&-0.174$\pm$ 0.024\\
 85&0.433$\pm$ 0.028&-0.037$\pm$ 0.002&-0.145$\pm$ 0.024\\
 90&0.456$\pm$ 0.028&-0.039$\pm$ 0.002&-0.121$\pm$ 0.023\\
 95&0.481$\pm$ 0.025&-0.042$\pm$ 0.002&-0.100$\pm$ 0.024\\
100&0.506$\pm$ 0.017&-0.045$\pm$ 0.003&-0.082$\pm$ 0.025\\ \hline
\end{tabular}
\end{center}
\end{table}

\begin{figure}
\begin{center}
\includegraphics[height=0.55\textheight,angle=0]{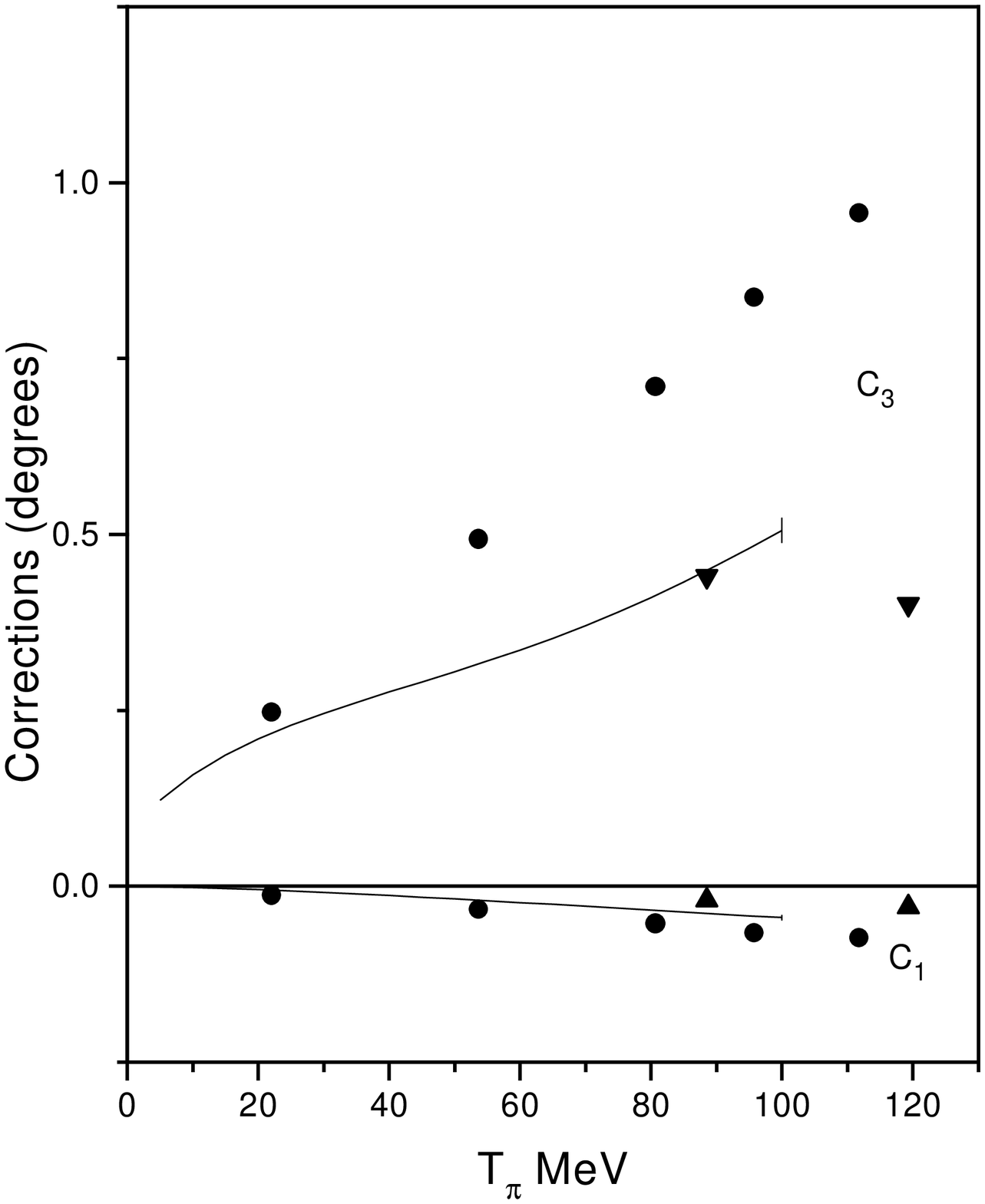}
\caption{Values in degrees of the electromagnetic corrections $C_{1}$ and
$C_{3}$ for the $p_{3/2}$-wave from our present calculation (solid curves),
from NORDITA \cite{2} (circles) and from Zimmermann \cite{9}
(triangles).}
\label{fig:4}
\end{center}
\end{figure}

\begin{figure}
\begin{center}
\includegraphics[height=0.55\textheight,angle=0]{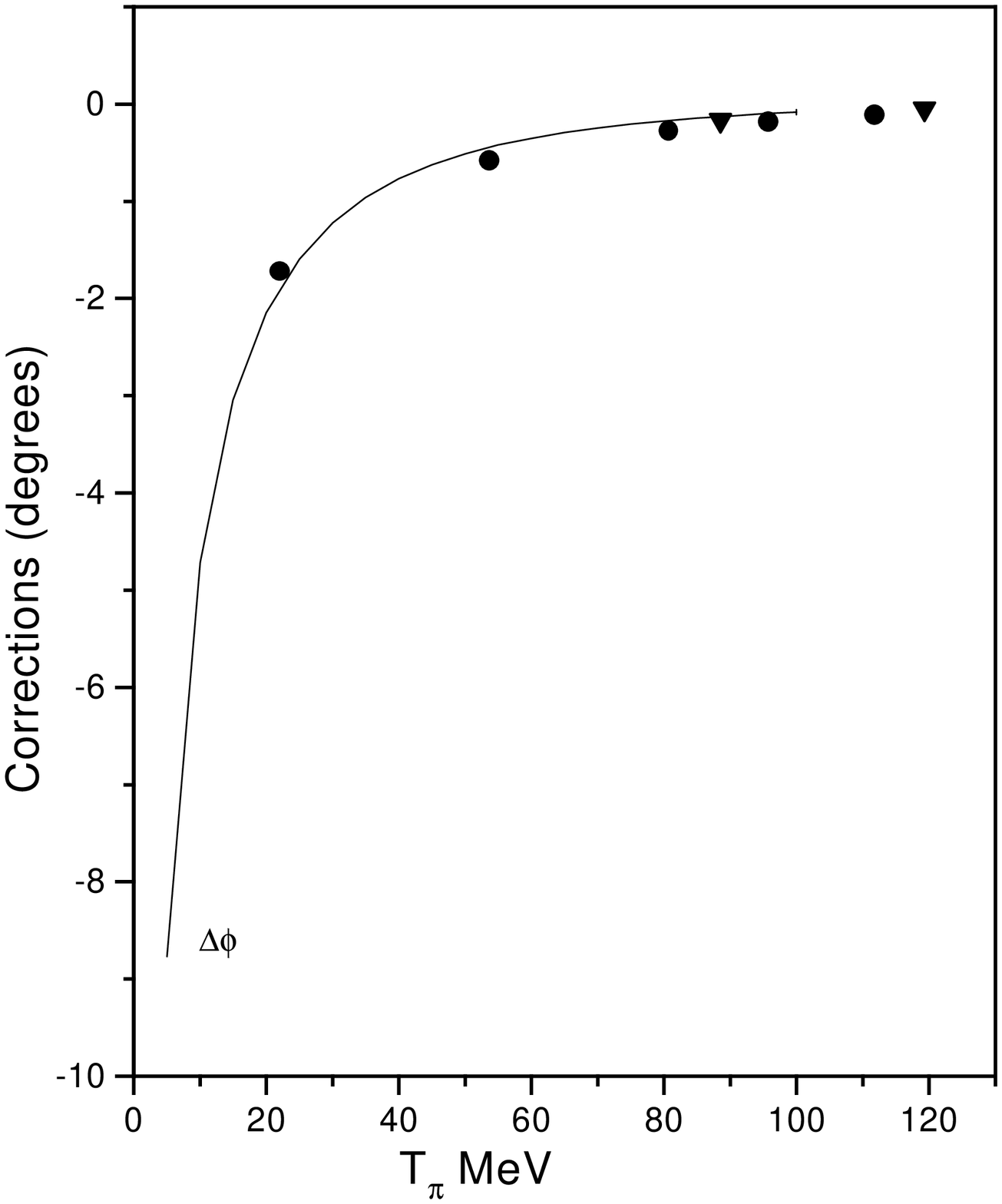}
\caption{Values in degrees of the electromagnetic correction $\Delta \phi$
for the $p_{3/2}$-wave from our present calculation (solid curve),
from NORDITA \cite{2} (circles) and from Zimmermann \cite{9}
(triangles).}
\label{fig:5}
\end{center}
\end{figure} 

\begin{table}
\begin{center}
\caption
{Values in degrees of the $p_{1/2}$-wave electromagnetic corrections $C_{3}$,
$C_{1}$ and $\Delta \phi$ as functions of the pion lab kinetic energy
$T_{\pi}$ (in MeV).}
\label{tab:3}
\begin{tabular}{|c|c|c|c|}
\hline
 $T_{\pi}$ & $C_{3}$ & $C_{1}$ &$\Delta \phi$ \\\hline
 10&-0.025$\pm$ 0.000&-0.042$\pm$ 0.001&15.911$\pm$ 0.052\\
 15&-0.031$\pm$ 0.001&-0.043$\pm$ 0.002&10.308$\pm$ 0.053\\
 20&-0.035$\pm$ 0.001&-0.042$\pm$ 0.003&7.288$\pm$ 0.052\\
 25&-0.037$\pm$ 0.001&-0.040$\pm$ 0.003&5.448$\pm$ 0.050\\
 30&-0.038$\pm$ 0.002&-0.037$\pm$ 0.003&4.225$\pm$ 0.048\\
 35&-0.039$\pm$ 0.002&-0.033$\pm$ 0.004&3.360$\pm$ 0.048\\
 40&-0.039$\pm$ 0.003&-0.028$\pm$ 0.004&2.720$\pm$ 0.050\\
 45&-0.038$\pm$ 0.003&-0.023$\pm$ 0.005&2.230$\pm$ 0.054\\
 50&-0.037$\pm$ 0.003&-0.017$\pm$ 0.005&1.844$\pm$ 0.063\\
 55&-0.036$\pm$ 0.004&-0.011$\pm$ 0.005&1.530$\pm$ 0.072\\
 60&-0.035$\pm$ 0.004&-0.005$\pm$ 0.005&1.268$\pm$ 0.083\\
 65&-0.034$\pm$ 0.004&0.002$\pm$ 0.005&1.032$\pm$ 0.102\\
 70&-0.032$\pm$ 0.004&0.009$\pm$ 0.006&0.910$\pm$ 0.105\\
 75&-0.031$\pm$ 0.005&0.015$\pm$ 0.005&0.851$\pm$ 0.108\\
 80&-0.030$\pm$ 0.007&0.023$\pm$ 0.004&0.784$\pm$ 0.110\\
 85&-0.029$\pm$ 0.007&0.030$\pm$ 0.004&0.689$\pm$ 0.107\\
 90&-0.027$\pm$ 0.007&0.037$\pm$ 0.004&0.572$\pm$ 0.108\\
 95&-0.026$\pm$ 0.008&0.043$\pm$ 0.005&0.483$\pm$ 0.109\\
100&-0.025$\pm$ 0.008&0.049$\pm$ 0.005&0.411$\pm$ 0.108\\ \hline
\end{tabular}
\end{center}
\end{table}

\section{Results for the corrections}

In Tables \ref{tab:1}, \ref{tab:2} and \ref{tab:3} we give the results for
the electromagnetic corrections, in the case of the $s$-, $p_{3/2}$- and
$p_{1/2}$-waves respectively, in the form of the corrections $C_1$ and $C_3$
to the hadronic phase shifts and the correction $\Delta \phi$ to the isospin
invariant mixing angle. They are given at 5 MeV intervals from $T_{\pi}=10$
MeV to $T_{\pi}=100$ MeV. The estimated uncertainties in the corrections are
also given in the tables. They were obtained in the same way as the
uncertainties in Table 1 of Ref. \cite{1}, by varying both the hadronic phase
shifts used as input and the range parameter in the hadronic potentials.  The
only case for which the uncertainty in $C_{3}$ or $C_{1}$ is comparable with
the error in the corresponding hadronic phase shift is $C_{3}$ for the
$p_{3/2}$-wave around 85 MeV, where the uncertainty in $C_{3}$ is
$0.028^{\circ}$ compared with an error in the phase shift of just over $0.03
^{\circ}$.

Inspection of the RSEs in Eq.(\ref{eq:21}) shows that the electromagnetic
corrections may be separated into two parts, one due to the inclusion of
$\mathbf{V}^{em}$ as an addition to $\mathbf{V}^h$ (Eq.(\ref{eq:20})) and the
other due to the difference between the masses of the particles in the two
channels ($q_c \neq q_0$, $m_c \neq m_0$, $f_c \neq f_0$).
The marked increase in  $\Delta \phi$ for small $T_{\pi}$ is due to the latter
effect. In Ref.\cite{1} we have also decomposed our results into the
contributions coming from the separate pieces of $V^{em}$ as given in
Eq.(\ref{eq:19}). Giving this decomposition in the coupled channel case would
involve us in complicated notation and would not convey any useful information:
the relative importance of the single pieces varies with the partial wave,
with the energy and with the particular correction ($C_1$, $C_3$ or $\Delta
\phi$).

Comparison with the results of the NORDITA group \cite{2} for the $s$- and
$p_{3/2}$-waves is complicated by the different quantities ($\Delta_1$,
$\Delta_3$, $\Delta_{13}$) that they use for the corrections. The relation
with our corrections is 
\[
\Delta_1=-\frac{3}{2}C_1  ,  \Delta_3=-3C_3  ,  \Delta_{13}=\frac{3}
{\sqrt{2}} \Delta \phi \sin(\delta_1^n-\delta_3^n)  ,
\]
the last relation being valid for $\Delta \phi$ very small. Our results are
compared with those of NORDITA \cite{2} and Zimmermann \cite{9} in Figs.
\ref{fig:2} and \ref{fig:3} for the s-wave and Figs. \ref{fig:4} and
\ref{fig:5} for the $p_{3/2}$-wave. We have indicated in Figs.
\ref{fig:2}-\ref{fig:5} the uncertainties in our corrections at $100$ MeV, as given
in Tables \ref{tab:1} and \ref{tab:2}. No errors are given in Refs. \cite{2}
and \cite{9} for the corrections presented there. The most important
differences are in 
$\Delta \phi$ for the s-wave and in $C_{3}$ for the $p_{3/2}$-wave. The former
is probably due to differences in the treatment of the mass differences, the
latter to the neglect in the NORDITA calculations  of medium and short range
effects due to $t$- and $u$-channel exchanges. The calculation of the $\pi
^{-}p$ corrections is not on as firm ground as the calculation of those for
$\pi^{+}p$. It is difficult to judge the treatment of the mass differences
in the NORDITA calculations because so little detail is given in Ref.
\cite{3}, while Ref. \cite{9} has a double energy dependence of the hadronic
potential term (a factor which is almost the same as our $\mathbf{f}$ at
low energies plus energy dependent potentials). We have remedied the obvious
deficiencies in these calculations and claim that as a result the values of
the corrections given in Tables \ref{tab:1} to \ref{tab:3} are more
reliable.

The corrections for the $p_{1/2}$-wave are given here for the first time. It
turns out that for the analysis of the data they play a negligible role, due
to the smallness of the hadronic phases in that partial wave.

The corrections in Tables \ref{tab:1}-\ref{tab:3} are intended for use in
future PSAs of $\pi^{-}p$ scattering experiments, provided that these PSAs
use Eqs.(\ref{eq:4}), (\ref{eq:5}), (\ref{eq:15}) and (\ref{eq:16}) and the
inelasticity corrections due to the $\gamma n$ channel given in
Eqs.(\ref{eq:13}) and (\ref{eq:14}). For very small angles or energies,
corrections to the expressions we have given for $f^{vp}$ and $\sigma_l^{vp}$
may also be needed, as discussed in Section 2 of Ref. \cite{1}.

\begin{ack} 
We thank the Swiss National Foundation and PSI (`Paul Scherrer Institut') for
financial support. We acknowledge very interesting discussions with W. R.
Gibbs and we are indebted to two referees for several very helpful comments
and suggestions.
\end{ack}


\begin{thebibliography}{99}
\bibitem{1} A. Gashi, E. Matsinos, G.C. Oades, G. Rasche and W.S. Woolcock,
preceding paper.
\bibitem{2} B. Tromborg, S. Waldenstr{\o}m and I. {\O}verb{\o}, Phys. Rev. D
15 (1977) 725. 
\bibitem{3} B. Tromborg , S. Waldenstr{\o}m and I. {\O}verb{\o}, Helv. Phys.
Acta 51 (1978) 584.
\bibitem{4a} W.R. Gibbs, Li Ai and W.B. Kaufmann, Phys. Rev. Lett.  74 (1995)
3740.
\bibitem{4} E. Matsinos, Phys. Rev.  C 56 (1997) 3014.
\bibitem{5} B.C. Pearce and B.K. Jennings, Nucl. Phys. A 528 (1991) 655.
\bibitem{6}  P.R. Auvil, Phys. Rev. D 4 (1971) 240.
\bibitem{7} W.R. Gibbs, Li Ai and W.B. Kaufmann, Phys. Rev. C 57 (1998) 784.
\bibitem{8} W.R. Gibbs, private communication. 
\bibitem{9} H. Zimmermann, Helv. Phys. Acta  48 (1975) 191.
\bibitem{10} G. Rasche and W.S. Woolcock, Helv. Phys. Acta  45 (1976) 495.
\bibitem{11} R.A. Arndt, R.L. Workman, I.I. Strakovsky and M.M. Pavan,
``$\pi N$ Elastic Scattering Analyses and Dispersion Relation Constraints'',
nucl-th/9807087; R.A. Arndt and L.D. Roper, SAID on-line program.
\end{thebibliography}
\end{document}